\documentclass{ws-procs975x65}

\def\generic{{\odot}}
\def\Rtt{{ R_{\hat t\hat t} }}
\def\Gtt{{ G_{\hat t\hat t} }}

\begin{document}


\wstoc{Generalized Puisieux series expansion for cosmological milestones}{C\'eline Catto\"en and Matt Visser}

\title{GENERALIZED PUISIEUX SERIES EXPANSION FOR COSMOLOGICAL MILESTONES}
\author{C\'eline Catto\"en and Matt Visser}

\address{School of Mathematics, Statistics, and Computer Science, \\
Victoria University of Wellington, \\
P.O.Box 600, Wellington, New Zealand\\
\email{celine.cattoen@mcs.vuw.ac.nz, matt.visser@mcs.vuw.ac.nz}}

\begin{abstract}
We use generalized Puisieux series expansions to determine the behaviour of the scale factor in the vicinity of typical cosmological milestones occurring in a FRW universe. We describe some of the consequences of this generalized Puisieux series expansion on other physical observables.
\end{abstract}

\bodymatter
\section{Introduction}
Over the last few years, the zoo of cosmological singularities considered in the literature has been considerably expanded, with ``big rips" and ``sudden singularities" added to the ``big bang'' and ``big crunch", as well as renewed interest in non-singular cosmological events such as ``bounces" and ``turnarounds"~\cite{rip, rip2, sudden2, sudden3, bounce}.

We consider a cosmological spacetime of the FRW form and assume applicability of the Einstein equations of general relativity. We will provide a generic definition of all the physically relevant singularities considered above (which we shall refer to as cosmological milestones), using generalized Puisieux series for the scale factor of the universe $a(t)$. We will show that, most importantly, \emph{all} physical observables ($H$, $q$,  the Riemann tensor, \emph{etc}...) will likewise be described by a generalized Puisieux series. 

\vskip-15pt

\section{Generalized Puisieux series expansion of the scale factor $a(t)$} 
Solutions of differential equations can often be expanded in Taylor series or Laurent series around their singular points. We shall extend this idea by expanding the scale factor $a(t)$ in generalized power series, similar to a Puisieux series, in the vicinity of the cosmological milestones.

\smallskip
\paragraph{Generic cosmological milestone:} 
Suppose we have some unspecified generic cosmological milestone,  that  is defined in terms of the behaviour of the scale factor $a(t)$, and which occurs at some finite time $t_\generic$. We will assume that in the vicinity of the milestone the scale factor has a (possibly one-sided) generalized power series expansion of the form
\begin{equation}
\label{E:power}
a(t) = c_0 |t-t_\generic|^{\eta_0} + c_1  |t-t_\generic|^{\eta_1} + c_2  |t-t_\generic|^{\eta_2} 
+ c_3 |t-t_\generic|^{\eta_3} +\dots
\end{equation}
where the indicial exponents $\eta_i$ are generically real (and are often non-integer) and without loss of generality are ordered in such a way that they satisfy
\begin{equation}
\eta_0<\eta_1<\eta_2<\eta_3\dots
\end{equation}
Finally we can also without loss of generality set $c_0 > 0$. There are no \emph{a priori} constraints on the signs of the other $c_i$, though by definition $c_i\neq0$.

The first term of the right hand side of equation (\ref{E:power}) is the dominant term, and is therefore responsible for the convergence or divergence of the scale factor at the time $t_\generic$. The indices $\eta_i$ are used to classify the cosmological milestones and the absolute value symbols are used to distinguish a past event from a future event. This generalized power series expansion of the scale factor is sufficient to represent almost all the physical models that we are aware of in the literature. Table \ref{table1} represents this cosmological milestone classification depending on the value of the scale factor.
\begin{table}[htdp]
\caption{Classification of cosmological milestones}
\begin{center}
\begin{tabular}{||c||c||c||}    \hline
Cosmological & Scale factor    & Indices  \\
milestones &  value                    & $\eta_i$ \\ \hline \hline
Big Bang/  &  $a(t_\generic)=0$   &  $\eta_0>0$ \\
Big Crunch&                                 &                  \\
 \hline
Sudden       & $a(t_\generic)=c_0$ & $\eta_0=0$ \\
Singularity   &$a^{(n)}(t_\generic)=\infty$& $\eta_1$ non-integer \\   \hline 
Extremality & $a(t_\generic)=c_0$  &  $\eta_0=0$ \\
events         &                                  & $\eta_i \in  \mathrm{Z^+}$ \\   \hline
Big rip           &$a(t_\generic)=\infty$ & $\eta_0<0$\\ \hline 
\end{tabular}
\end{center}
\label{table1}
\end{table}
Note that sudden singularities are of order $n$ where the $n^{th}$ derivative of  the scale factor is the first one that is infinite:
\begin{equation}
a^{(n)}(t\to t_\generic) \sim c_0 \; \eta_1 (\eta_1-1)(\eta_1-2)\dots (\eta_1-n+1) \; |t-t_\generic|^{\eta_1-n}\to\infty,
\end{equation}
and therefore $\eta_1$ has to be a non-integer~\cite{sudden2, sudden3}.
Note that for most calculations it is sufficient to use the first three (or fewer) terms of the power series expansion.

\section{Power series expansion of all physical observables} 
We have exhibited a generic expansion of the scale factor $a(t)$ based on generalized power series  for all the physically relevant cosmological milestones found in the literature to date (big bang, big crunch, sudden singularity, extremality events and big rip). We can now use the parameters of this series to explore the kinematical and dynamical properties of the cosmological milestones, for example, to see whether they are true curvature singularities or whether the energy conditions hold in the vicinity of the time of the event $t_\generic$.

For instance, on a kinematical level, we can analyze the Hubble parameter for finiteness in the vicinity of the cosmological milestones. Keeping the most dominant terms, we have for $\eta_0\neq0$:
\begin{equation}
H ={\dot a\over a} 
\sim 
{c_0 \eta_0 (t-t_\generic)^{\eta_0-1}\over c_0 (t-t_\generic)^{\eta_0}} 
=  
{\eta_0\over t-t_\generic};  \qquad (\eta_0\neq0).
\end{equation}
That is, for bangs, crunches, and rips the Hubble parameter exhibits a generic $1/(t-t_\generic)$ blow up.

In a similar fashion, we can also determine whether a cosmological milestone is a \emph{true curvature singularity} by testing $\Rtt$ and $\Gtt$ in orthonormal components for finiteness:
\begin{equation}
\Rtt = - 3 \;{\ddot a\over a}; \qquad
\Gtt = 3 \left( {\dot a^2\over a^2} + {k\over a^2}\right).
\end{equation}

On a dynamical level, we can quantify how ``strange" physics gets in the vicinity of a cosmological milestone by introducing the Friedmann equations and the standard energy conditions in general relativity --- which are the \emph{null}, \emph{weak}, \emph{strong}, and \emph{dominant} energy conditions~\cite{bounce, Tolman, twilight}. The density and pressure are given as a function of the scale factor $a(t)$ and can therefore likewise be power series. Whether or not a specific energy condition is satisfied is simply a matter of calculating the dominant indicial exponents of the series expansion (full details provided in~\cite{Cattoen:2005dx}).

To conclude, if in the vicinity of any cosmological milestone, the input scale factor $a(t)$ is a generalized power series, then all physical observables (\emph{e.g.} $H$, $q$, the Riemann tensor, \emph{etc}.) will likewise be a generalized Puisieux series. By checking the related indicial exponents, which can be calculated from the indicial exponents of the scale factor, one can determine whether or not the particular physical observable then diverges at the cosmological milestone.


\enlargethispage{25pt}

\end{document}